# Influence of Pb addition on the superconducting properties of polycrystalline $Sr_{0.6}K_{0.4}Fe_2As_2$

Lei Wang, Yanpeng Qi, Zhiyu Zhang, Dongliang Wang, Xianping Zhang, Zhaoshun Gao, Chao Yao, Yanwei Ma[*]

Key Laboratory of Applied Superconductivity, Institute of Electrical Engineering, Chinese Academy of Sciences, P. O. Box 2703, Beijing 100190, China

**Abstract**

Polycrystalline $Sr_{0.6}K_{0.4}Fe_2As_2$ samples with various Pb additions (0-20 wt%) were prepared using a one-step solid state reaction. X-ray diffraction analysis shows no evidence for chemical reaction between the Pb and the FeAs-based superconductor. However, the presence of the Pb can affect the microstructure and superconducting properties of the final products. The critical transition temperature $T_c$ indicates no degradation up to 20 wt% Pb addition, and dramatic improvements of magnetic $J_c$ and irreversibility field $H_{irr}$ were observed for appropriate Pb concentration. Transport critical current property of pure and Pb-added $Sr_{0.6}K_{0.4}Fe_2As_2$ tapes was also measured by a four-probe technique, and a remarkable enhancement of $J_c$ at low fields was detected for the Pb added tapes.

---

[*] Author to whom correspondence should be addressed; E-mail: ywma@mail.iee.ac.cn

**Introduction**

The recent discovery of superconductivity at 26 K in the LaFeAsO$_{1-x}$F$_x$ compound [1] has generated tremendous interests among physicists and material scientists [2-4]. Further research has led to the discovery of superconductivity at 38 K in Ba$_{0.6}$K$_{0.4}$Fe$_2$As$_2$ [5]. The parent compounds, REFeAsO (1111 type, RE-rare earth) and AFe$_2$As$_2$ (122 type, A=Ba, Sr), have a quasi two-dimensional tetragonal structure, which consist of charged (LaO)$^{\delta+}$ layers or A$^{\delta+}$ alternating with (FeAs)$^{\delta-}$ layers. Superconductivity was produced by doping the parent compounds with electrons or holes, and the highest Tc that has been found in this FeAs-based system is around 55 K [6]. It has been accepted that these materials represent the second class of high-$T_c$ superconductors after the discovery of cuprates in 1986 [7].

The K doped AFe$_2$As$_2$ (122 type, A=Ba, Sr) exhibits a high critical transition temperature of ~38 K, together with high critical current density $J_c$, low anisotropy, and high critical fields [8-10]. On the other hand, the synthesizing temperature is relatively low (~850°C) and no oxygen is involved, compared with that of the RE-1111 series. All these would be advantages for potential applications [8-10]. Recently, transport critical current density $J_c$ of 1.2×10$^3$ A/cm$^2$ at 4.2 K in self field has been observed in Sr$_{0.6}$K$_{0.4}$Fe$_2$As$_2$ / Ag tapes [11]. This result would give further encouragement to the development of the newly discovered FeAs-based superconductors for potential high current applications.

Generally, chemical addition usually plays an important role in enhancing superconducting properties, by promoting the crystallization of the superconducting phase, catalyzing the intergranular coupling of the superconducting grains or introducing pinning centers. For example, the irreversibility field H$_{irr}$ can be largely increased by C addition in MgB$_2$, and a $J_c$ enhancement was observed in Ag added YBCO [12-14]. Recently, we have reported that the H$_{irr}$ and superconducting properties of the 122 phase iron-based superconductor can be significantly increased by Ag addition [11, 15]. On the other hand, it has been found that Pb addition effectively improved superconducting properties of BSCCO, because of a rearrangement of the band structure, as well as facilitating grain growth [16-17]. In this paper, we studied the effect of Pb on the microstructure and superconducting properties of polycrystalline $Sr_{0.6}K_{0.4}Fe_2As_2$, and found that an enhancement of $J_c$ in pnictide bulks and tapes can be achieved by Pb addition.

**Experimental details**

The polycrystalline $Sr_{0.6}K_{0.4}Fe_2As_2$ investigated were prepared by a one-step solid-state reaction method developed by our group [18], together with ball milling process. Sr filings, Fe powder, As and K pieces, with a ratio Sr : K : Fe : As = 0.6 : 0.4 : 2 : 2, were thoroughly ground in Ar atmosphere for more than 10 hours using ball milling method. After the ball milling, Pb were added into the raw powder, and then the mixture were ground in a mortar for half an hour. The final powders were filled and sealed into Nb tubes (OD: 8 mm, ID: 5 mm), the Nb tube was subsequently rotary swaged. The samples were sintered at 500°C for 15 hours, and then at 850-900°C for 35 hours in Ar atmosphere. Four kinds of samples with various Pb additions (0, 5, 10 and 20 wt%) were made for this study. The density of the sintered $Sr_{0.6}K_{0.4}Fe_2As_2$ sample is about 70% of the theoretical value of 5.89 g·cm$^{-3}$. The fabrication of $Sr_{0.6}K_{0.4}Fe_2As_2$ tapes were described in previous papers [11, 18].

Phase identification was characterized by powder X-ray diffraction (XRD) analysis with Cu-Kα radiation from 20 to 80°. Resistivity measurements were carried out by the standard four-probe method using a PPMS system. DC susceptibility of the $Sr_{0.6}K_{0.4}Fe_2As_2$ samples was measured using a Quantum Design SQUID. The samples were first cooled down to 5 K in zero magnetic field and then a magnetic field of 20 Oe was applied. The diamagnetic susceptibility due to the shielding current was measured during the warming process up to the temperatures well above $T_c$ (ZFC: zero-field cooled). In a field of 20 Oe, the Meissner effect was measured during the cooling process. Rectangular specimens in the dimension of about 5×2.5×1.5 mm$^3$ were cut from the samples, and magnetization measurements were performed with a PPMS system in fields up to 7 T. Magnetic critical current densities were calculated using Bean model ***Jc=20Δm/va(1-a/3b)***, taking the full sample dimensions. Microstructural observations were performed using scanning electron microscope (SEM). The transport critical currents of the pure and doped $Sr_{0.6}K_{0.4}Fe_2As_2$ tapes at 4.2 K and its magnetic dependence were evaluated at the High Field Laboratory for Superconducting Materials (HFLSM) in Sendai, Japan, by a standard four-probe resistive method, with a criterion of 1 $\mu$V cm$^{-1}$.

**Results and discussion**

The powder X-ray diffraction patterns of the pure and Pb added $Sr_{0.6}K_{0.4}Fe_2As_2$ bulk samples are shown in Fig. 1. The pattern of pure sample is well indexed to $Sr_{0.6}K_{0.4}Fe_2As_2$. No obvious foreign phase was detected, ensuring that the proposed treatment was successful to obtain a single phase $Sr_{0.6}K_{0.4}Fe_2As_2$ sample. The Pb added samples consist of $Sr_{0.6}K_{0.4}Fe_2As_2$ as the major phase, however, Pb peaks were clearly detected in the Pb added samples, and a small amount of FeAs phase was also identified, particularly in the nominal 10 % and 20 % Pb added samples.

Figure 2 presents normalized resistivity versus temperature curves for $Sr_{0.6}K_{0.4}Fe_2As_2$ bulk samples with various Pb concentrations. All samples exhibit sharp resistive superconducting transition at $T_c$ (onset) ≈ 35 K with $\Delta T_c \leq 2.5$ K. In particular, the 5 % Pb added samples show a zero transition at ~34 K. There is a general tendency for normalized resistivity above $T_c$ to decrease with Pb addition, and the residual resistivity ratio RRR = $\rho(300K)/\rho(40K)$ for the pure and Pb added samples are 3.8 and 4.4, respectively. The resistive data suggest that the $T_c$ was hardly affected by Pb addition

DC susceptibilities of pure and Pb added $Sr_{0.6}K_{0.4}Fe_2As_2$ bulk samples were measured, and all the results are shown in Fig. 3. The ZFC curve for pure samples indicates that the significant shielding currents appear at about 30 K, and increases as temperature decreases. However, the shielding currents for the Pb added samples occur at 35 K (Point A), which may suggest a crystallization improvement caused by Pb addition. The shielding currents for Pb were also seen at 9 K (Point B), which increase steadily with increasing the nominal Pb content. It is also found that, the meissner response for the Pb added samples is stronger than that of the pure samples, indicative of larger meissner fraction in Pb added samples.

Shown in the figure 4 are the magnetic $J_c$ (at 5 K) vs magnetic field curves for the Pb added and pure bulks. As is evident from the figure, magnetic $J_c$ in the entire field region can be increased by Pb addition. A substantial improvement is obtained by increasing the Pb content up to 10 %, while upon further increasing the Pb content, the magnetic $J_c$ decreases, because of too much non-superconducting phase existed (see Fig. 1). The $J_c$ of 5 % Pb added samples at 5 K in self field is about $1.5 \times 10^4$ A/cm$^2$ and still remains above $1 \times 10^3$ A/cm$^2$ beyond 6.5 T, twice as high as for the pure samples. Most importantly, large $J_c$ of about $2 \times 10^4$ A/cm$^2$ at 5 K in self field was achieved in the 10 % Pb added samples.

The $J_c$ of pure and 5 % Pb added samples, as a function of magnetic fields at various temperatures, is given in the inset of Fig. 4. As expected, significant improvement of $J_c$ at 10 K, 20 K and 30 K induced by Pb addition can be achieved. To our surprise, the $J_c$ of the 5 % Pb added sample maintains ~$10^2$ A/cm$^2$ even at 30 K and high magnetic fields. Although a remarkably enhanced magnetic $J_c$ was observed in Pb added samples, it is known that, two kinds of loops, intra-grain current loops and inter-grain current loops, contribute to the magnetization of the bulk sample. Thus, the improvement in magnetic $J_c$ may originate from intra-grain current loops, or inter-grain current loops, or both of them.

Figure 5 depicts the variation of resistive transitions under various magnetic fields (H=0, 1, 3, 5, 7 and 9 T) for the pure and a 5 % Pb added samples. The resistive transition regions clearly broaden as the magnetic field increases, with only a small effect in the region near the onset of transitions. One striking feature here is that the zero transition of Pb added samples is much less sensitive to magnetic field than that of the pure samples. Upper critical field $H_{c2}$ ($T$) (Filled squares for the pure samples, and filled circles for 5 % Pb added samples) and irreversibility field $H_{irr}$ ($T$) (Open squares for the pure samples, and open circles for 5% Pb added samples) were estimated, using criteria of 90% and 10 % of normal state resistivity respectively, as shown in the inset of Fig. 5. Note that $H_{c2}$ ($T$) was not significantly changed by Pb addition. Upper critical field was extrapolated to 0 K using the Werthamer-Helfand-Hohenberg (WHH) formula, $H_{c2}$ (0) = - 0.693 $T_c$ (d$H_{c2}$/d$T$). The slope d$H_{c2}$/d$T$ estimated from the $H$-$T$ phase diagram is about 8 for both samples. Taking $T_c$ = 34 K, the upper critical field is $H_{c2}$ (0) = 188 T. In contrast, the addition of Pb produces a large enhancement of irreversibility field in $Sr_{0.6}K_{0.4}Fe_2As_2$. For instance, the $H_{irr}$ for the pure samples is 3 T at 30 K, however, the extrapolated $H_{irr}$ for the 5 % Pb added sample at 30 K is about 18 T, 15 T higher than that of the pure samples.

To further understand the Pb effects, the microstructure of pure and Pb added samples was studied as shown in Fig.6. As we can see, a dense structure with a grain size of about 5 μm in average diameter is observed in Pb added samples (Fig. 6b), much larger than the size of grains (~1 μm) in the pure samples (Fig. 6a), indicating a substantial grain enlargement, which was supposed to occur during the sintering process by Pb vapor or liquid. EDX analysis on a large area in the Pb added samples clearly demonstrates that the product is composed of Sr, K, Fe, As and Pb elements and no other impurity element was found (Fig. 6c). In addition, SEM investigation using quadrant back scattering detector (Fig. 6d) shows small Pb particles dispersing in parent compound, with some residing between grains.

Therefore, it can be concluded that Pb addition inherently modifies grain dimension as well as accelerates growth, which may be responsible for some positive effects, such as the sharp resistive transition in 5 % Pb added samples and enhanced irreversibility field and magnetic $J_c$. However, heavy Pb addition results in a Jc degradation, because of more non-superconducting phases existed, which is supported by XRD analysis in the 10 % and 20% Pb added samples (see Fig.1).

In order to reveal the effect of Pb addition on the transport property, some pure and Pb added $Sr_{0.6}K_{0.4}Fe_2As_2$ tapes were made through the in-situ powder-in-tube method, and transport critical currents of the tapes were measured by using a standard dc four-probe method, as shown in Fig. 7. Clearly, the Pb added tapes show a higher $J_c$ than the pure samples in low field region, and a highest $J_c$ of 1100 A/cm$^2$ in self field was obtained by 5 wt% Pb addition. However, the $J_c$ in high field region was not significantly increased. The preparation and details of superconducting properties of the Pb added wires and tapes will be reported elsewhere.

Although a remarkably enhanced magnetic $J_c$ was observed in Pb added samples, it is known that, two kinds of loops, intra-grain current loops and inter-grain current loops, contribute to the magnetization of granular superconductors. Thus, the improvement in magnetic $J_c$ (Fig. 4) may originate from intra-grain current loops, or inter-grain current loops, or both of them. As the SEM study on microstructures (Fig. 6) reveals that Pb additions promote crystal growth, these large grains, meaning large dimensions of intra-grain loops, were supposed to contribute to the enhancement of magnetic $J_c$ in the entire field region. In addition, the transport result (Fig.7) clearly shows that another contribution from inter-grain $J_c$ improvement also exists.

Therefore, the improved magnetic $J_c$ originates from both intra-grain and inter-grain currents, then the enlarged intra-grain current loops, maybe due to the grain size enlargement by Pb addition, are responsible for enhanced magnetic $J_c$ at high fields. While for the improvement of transport $J_c$ at low fields, further study on grain boundary is needed.

**Conclusions**

We have demonstrated effects of Pb additions on critical transition temperature $T_c$, magnetic hysteresis, upper critical field $H_{c2}$, irreversibility field $H_{irr}$, and transport $J_c$ of polycrystalline $Sr_{0.6}K_{0.4}FeAs$. The critical transition temperature $T_c$ showed no degradation up to 20 wt% Pb addition, and dramatic enhancements of magnetic $J_c$ and irreversibility field $H_{irr}$ were observed for appropriate Pb concentration. We notice that even transport $J_c$ is not affected much at high fields, a substantial improvement is obtained at low fields by 5 wt% Pb addition.

**Acknowledgements**

The authors thank Profs. K. Watanabe, S. Awaji, G. Nishijima, Haihu Wen and Liye Xiao for their help and useful discussions. This work is partially supported by the Beijing Municipal Science and Technology Commission under Grant No. Z09010300820907, National '973' Program (Grant No. 2006CB601004) and Natural Science Foundation of China (Grant No: 50777062 and 50802093).

**Captions**

Figure 1 X-ray diffraction patterns of pure and Pb added $Sr_{0.6}K_{0.4}Fe_2As_2$ samples.

Figure 2 Normalized resistivities for pure and Pb added samples as a function of temperature. Inset: Temperature dependence of normalized resistivities near critical transition.

Figure 3 Temperature dependence of DC susceptibility for pure and Pb added $Sr_{0.6}K_{0.4}Fe_2As_2$ samples.

Figure 4 Variation of magnetic $J_c$ as a function of applied field for pure and Pb added $Sr_{0.6}K_{0.4}Fe_2As_2$ bulk samples at 5 K; Inset: The magnetic $J_c$ of pure and 5 % Pb added $Sr_{0.6}K_{0.4}Fe_2As_2$ samples at various temperatures.

Figure 5 Temperature dependence of resistivity for pure (black squares) and 5 % Pb added (red circles) $Sr_{0.6}K_{0.4}Fe_2As_2$ samples in various fields H=0, 1, 3, 5, 7 and 9 T. Inset: Phase diagram of pure (black squares) and 5 % Pb added (red circles) $Sr_{0.6}K_{0.4}Fe_2As_2$ samples as determined from field dependence of critical transition.

Figure 6 Scanning electron micrographs of pure (a) and 10 % Pb added (b) $Sr_{0.6}K_{0.4}Fe_2As_2$ samples; EDX spectrum (c) and QBSD image (d) for the 10 % Pb added samples.

Figure 7 Transport $J_c$ at 4.2 K as a function of applied field for pure and Pb added $Sr_{0.6}K_{0.4}Fe_2As_2$ tapes. The measurements were performed in magnetic fields parallel to the tape surface.

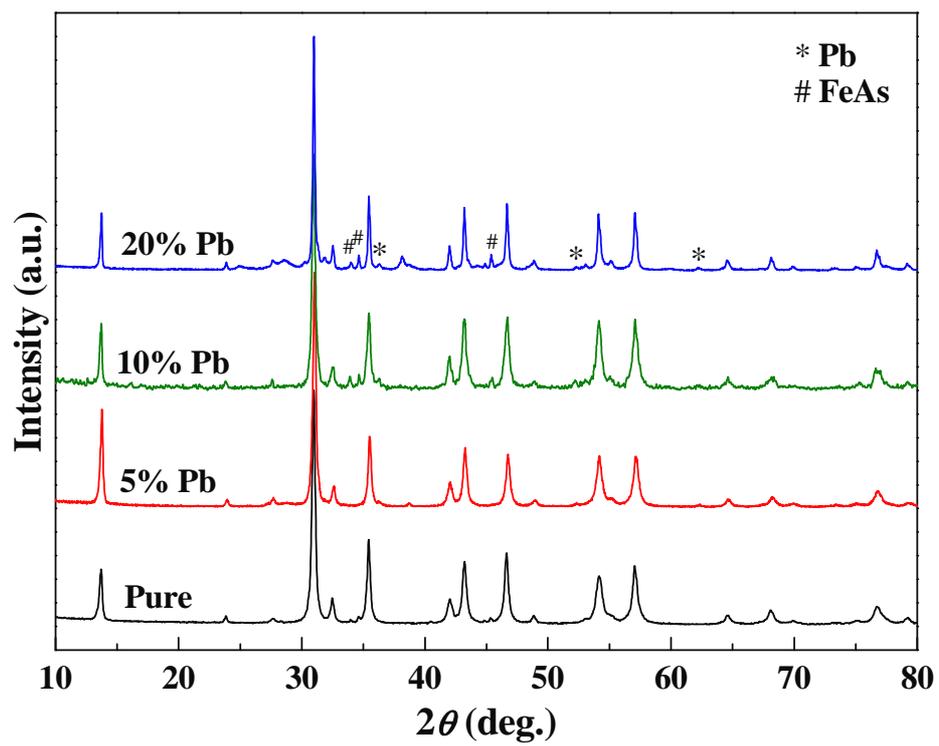

Figure 1 Wang et al.

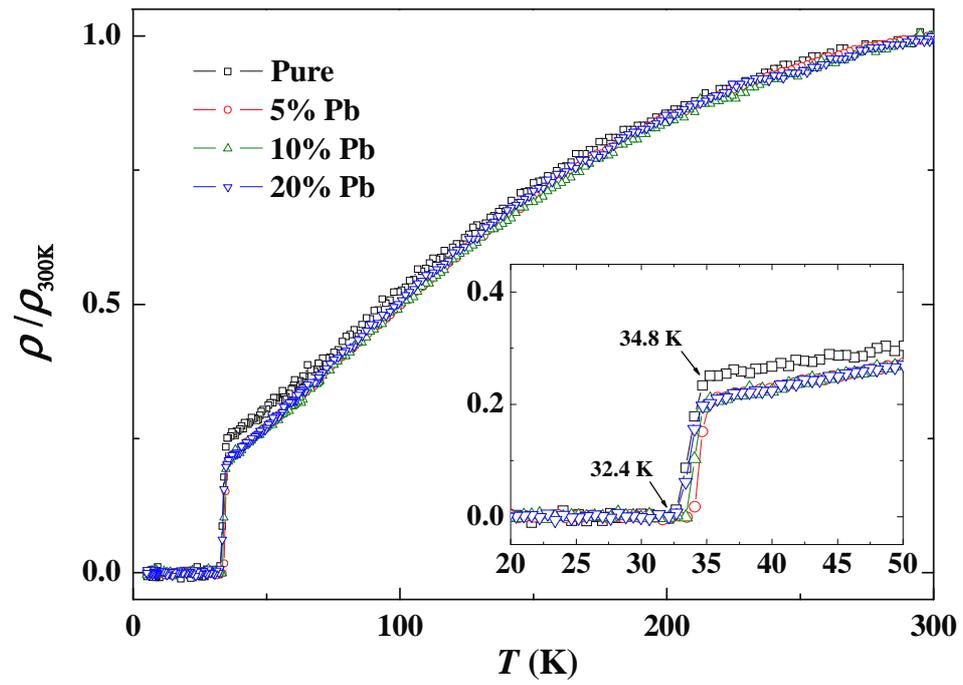

Figure 2 Wang et al.

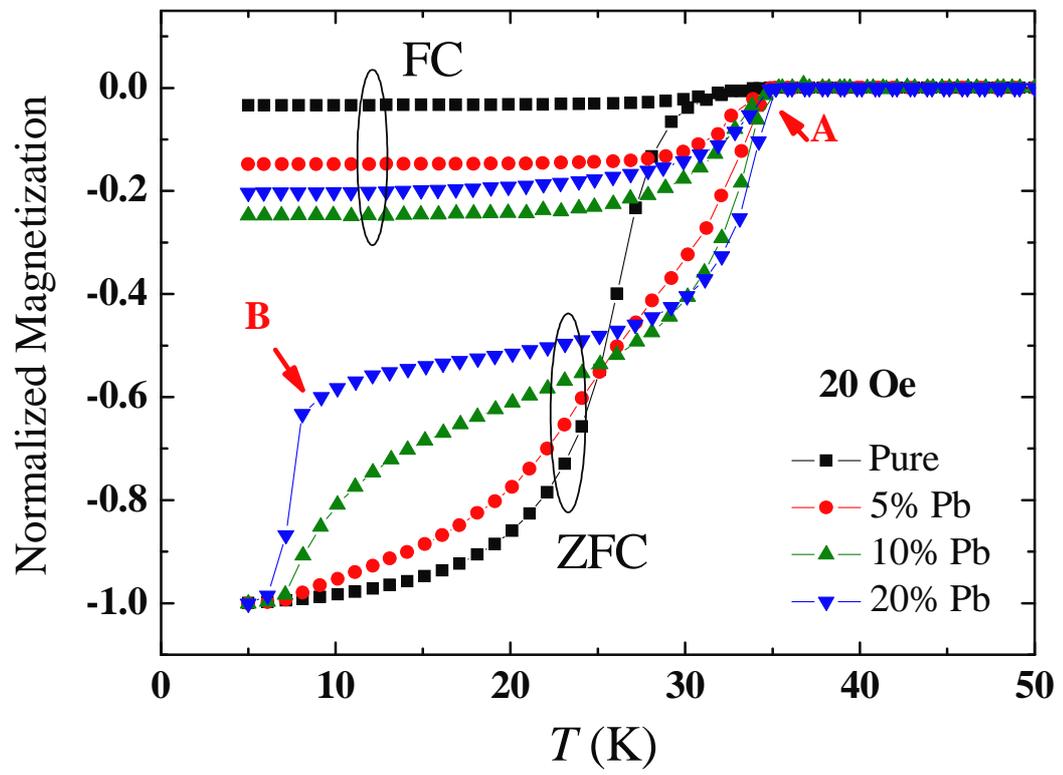

Figure 3 Wang et al.

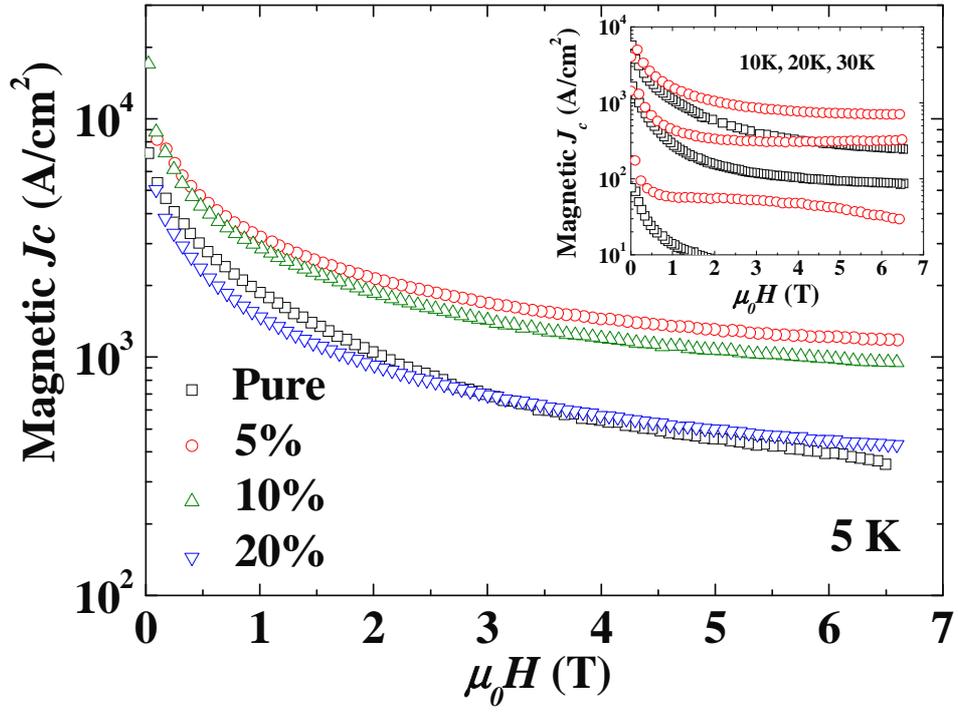

Figure 4 Wang et al.

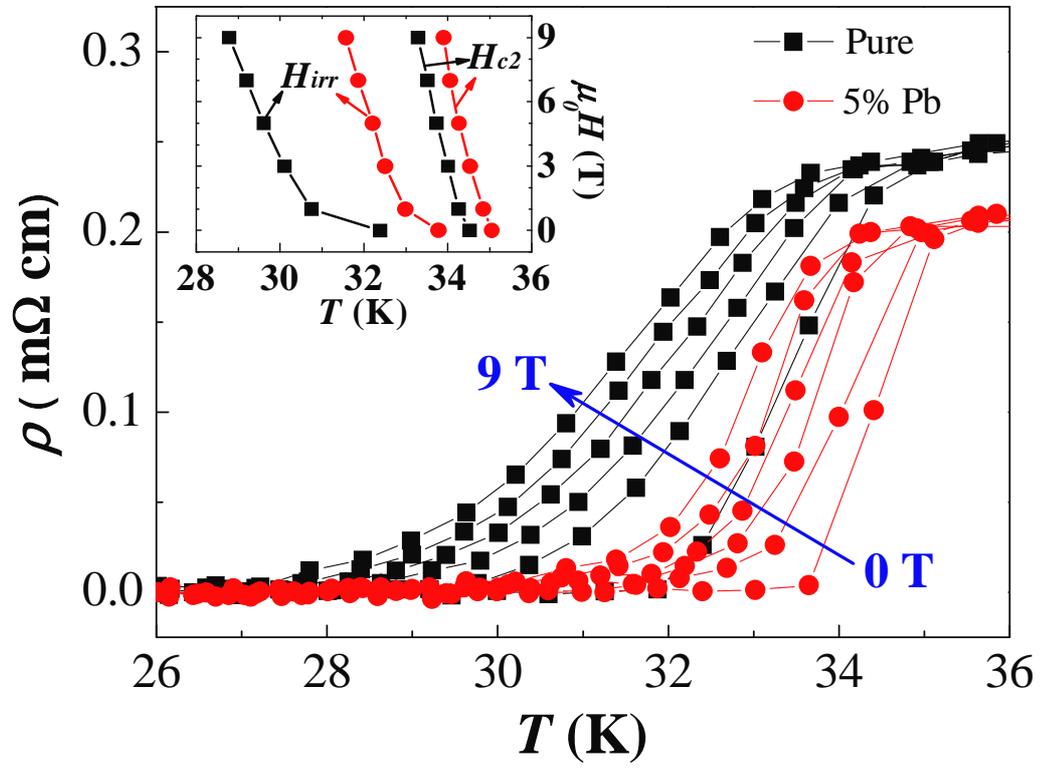

Figure 5 Wang et al.

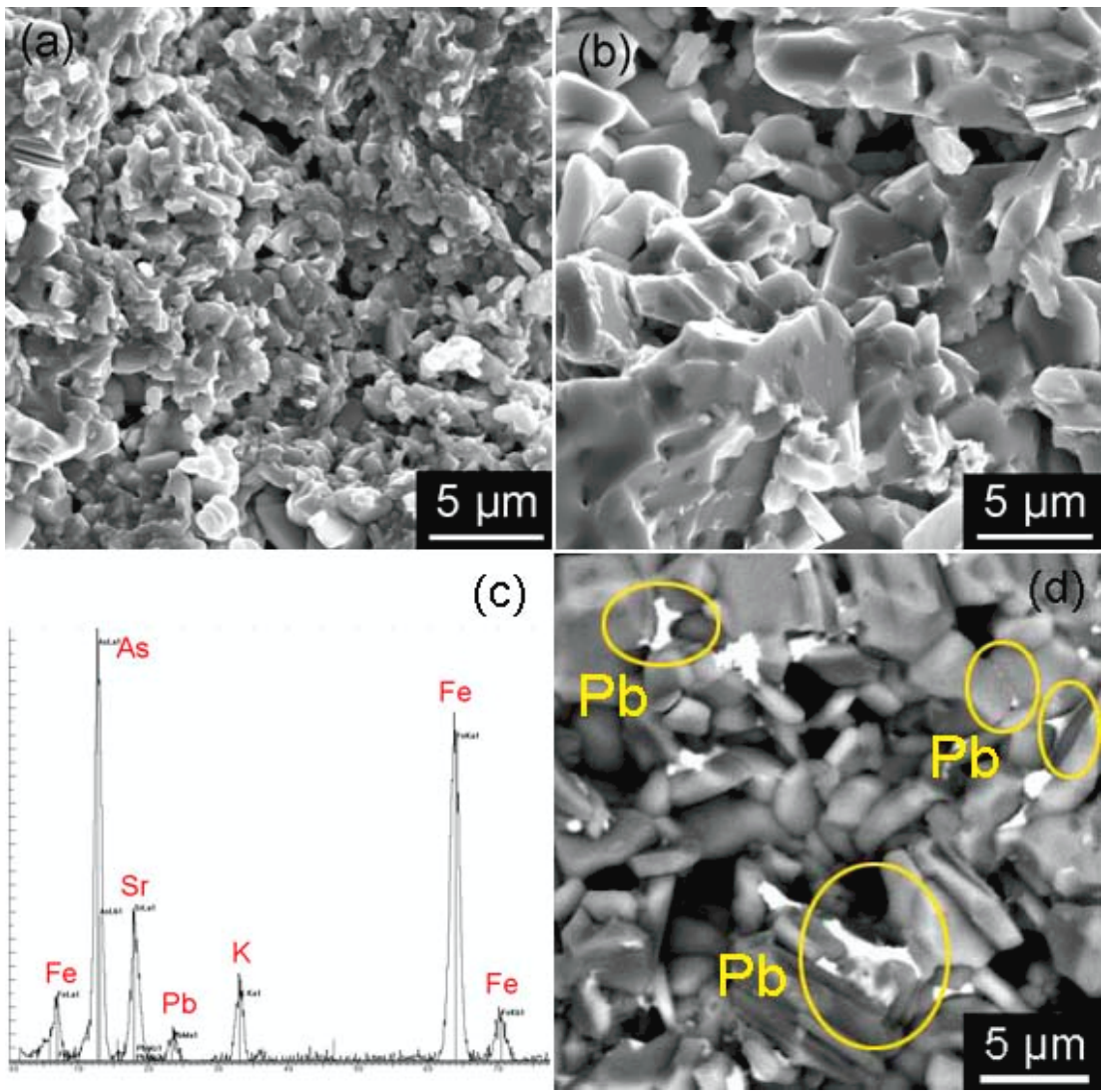

Figure 6 Wang et al.

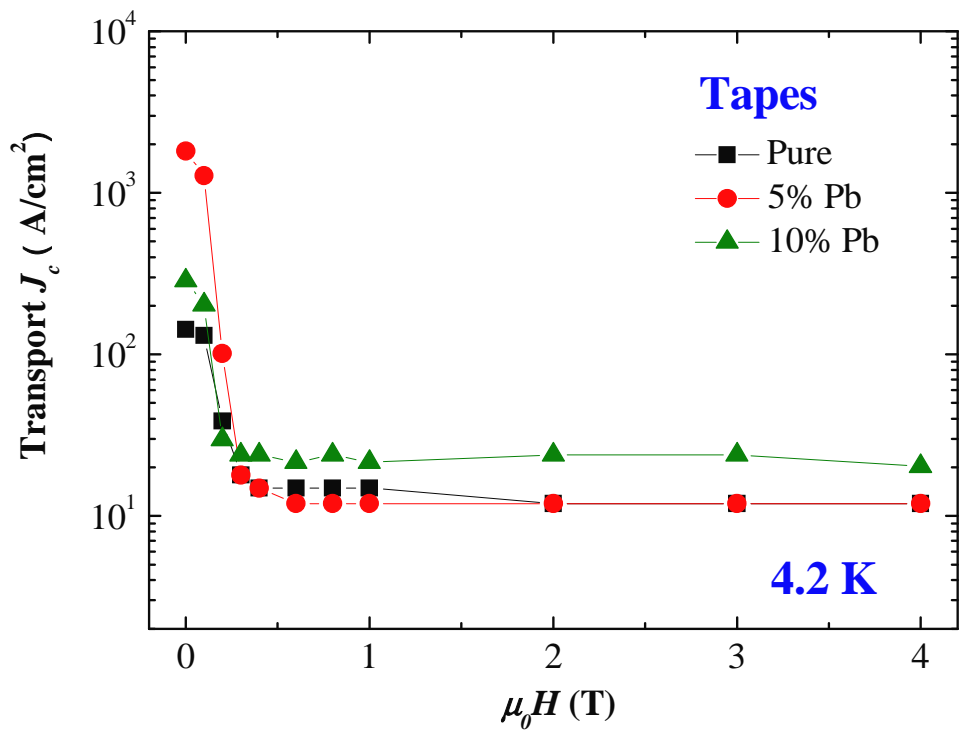

Figure 7 Wang et al.